# Effect of Fluoropolymer Composition on Topochemical Synthesis of $SrMnO_{3-\delta}F_{\gamma}$ Oxyfluoride Films

*Jiayi Wang,[1] Yongjin Shin,[2] Elke Arenholz,[3] Benjamin M. Lefler,[1] James M. Rondinelli,[2]*

*Steven J. May[1,*]*

[1]Department of Materials Science and Engineering, Drexel University, Philadelphia,

Pennsylvania 19104, USA

[2]Department of Materials Science and Engineering, Northwestern University, Evanston, Illinois

60208, USA

[3]Advanced Light Source, Lawrence Berkeley National Laboratory, One Cyclotron Road,

Berkeley, California 94720, USA

* smay@drexel.edu

We report the synthesis of $SrMnO_{3-\delta}F_{\gamma}$ perovskite oxyfluoride thin films using a vapor transport method to fluorinate as-grown $SrMnO_{2.5}$ epitaxial thin films. The influence of the fluoropolymer, which acts as a fluorine vapor source, was investigated by utilizing polyvinyl fluoride (PVF), polyvinylidene difluoride (PVDF) and polytetrafluoroethylene (PTFE) in the reaction. The same process was carried out with polyethylene (PE) to isolate the role of carbon in the vapor transport process. The F distribution was probed by X-ray photoemission spectroscopy, which confirmed the incorporation of F into the films and revealed higher F concentrations in films exposed to PVF and PVDF compared to PTFE. The *c*-axis parameter expands after fluorination, a result consistent with density functional theory calculations that attribute the volume expansion to elongated Mn-F bonds compared to shorter Mn-O bonds. Using X-ray absorption spectroscopy, we show that the fluorination process reduces the nominal Mn oxidation state suggesting that F substitutes on O sites in the lattice as opposed to filling anion vacancy sites, a finding further supported by calculated formation energies of different F site occupancies. These results provide new insights into topochemical fluorination of perovskite oxides, which should enable future synthesis and design efforts focused on oxyfluoride heterostructures.

**I. INTRODUCTION**

Combining multiple chemically distinct anions into functional materials presents a useful route to tune and design material properties [1,2], ranging from superconductivity [3-5] to ionic conductivity [6,7]. In particular, the metal oxyfluorides are of interest as incorporation of both $F^-$ and $O^{2-}$ provides a means to alter metal-anion bond ionicity and the nominal metal oxidation state. For example, if F replaces O through anionic substitution, the transition metal will be reduced, increasing its nominal electron count. The use of F as an electron dopant has been employed in doping cuprates into the superconducting state [8] and in reducing Fe to a 3+ oxidation state through conversion of $SrFeO_3$ to $SrFeO_2F$ [9]. In $NdNiO_3$, an orders of magnitude increase in room temperature resistivity was observed upon fluorination, highlighting the significant electronic modifications that can be induced through fluorine incorporation on the anion site [10]. Additionally, structural changes induced through fluorination to metal-anion bonding environments have also been reported to play a central role in superconductivity in cuprates [3]. In perovskite oxides, the incorporation of F on the anion site is more accessible compared to other halides because $F^-$ and $O^{2-}$ are relatively similar in size [1], often allowing for the crystal structure to be maintained upon fluorination, which may not be feasible with incorporation of larger halide ions such as $Cl^-$ or $Br^-$. The limit to the F concentration in perovskite oxyfluorides is typically set by the oxidation states that the *B*-site cation can accommodate [11]. Early synthesis efforts of transition metal oxyfluorides were carried out by solid-state reactions at around 1000°C [12]. In order to reduce the synthesis temperature, topochemical fluorination of metal oxide precursors with fluorine sources including $F_2$, $NH_4F$, $MF_2$ ($M$ = Ba, Cu, Ni, Zn), and $XeF_2$ was utilized, significantly reducing the reaction temperature [5]. This fluorination method has been reported in producing powder samples of copper, titanium, iron, manganese and other metal oxyfluorides [3-6,13].

An alternative approach to fluorination at low temperatures was introduced by Slater [14], who demonstrated that polyvinylidene difluoride (PVDF) can be used as a fluorine vapor source when decomposed in close proximity to metal oxide powders. Following the work of Slater, PVDF has been applied in fluorination of other metal oxide samples in powder form, especially in producing the perovskite-related oxyfluoride materials [9,15-17]. The use of fluoropolymers can further reduce the fluorination temperature to 180°C for PVDF and 330°C for polytetrafluoroethylene (PTFE) [11], and mitigate the formation of secondary phases. Additionally, polymer-based fluorination has been applied to the synthesis of oxyfluoride thin films, carried out as post-growth reaction on as-grown films [18-21]. To date, these fluorination studies of thin films have all utilized PVDF as the fluorine source, while PTFE has been only investigated in producing bulk oxyfluorides [22]. To the best of our knowledge, polyvinyl fluoride (PVF) has not been reported as a fluorinating agent yet. Thus, there has been no systematic report of how the choice of fluoropolymer influences the fluorination reaction and resultant oxyfluoride films.

We have synthesized epitaxial $SrMnO_{3-\delta}F_\gamma$ (SMOF) oxyfluoride thin films by fluorinating the as-grown $SrMnO_{2.5}$ (SMO) films with PVF, PVDF, and PTFE. We find that the use of PVDF and PVF results in more F incorporation than PTFE; however, the crystalline quality decreases with increasing F content. The c-axis parameters are observed to expand with the F incorporation, consistent with the formation of Mn-F bonds. After fluorination, the nominal 3+ valence state of Mn in the as-grown film was reduced toward 2+ suggesting that F is incorporated through substitution for O as opposed to insertion into anion vacancy sites.

## II. METHODS

Epitaxial films of $SrMnO_{3-\delta}$ were deposited by oxide molecular beam epitaxy (MBE) on $10\times10$ mm$^2$ $(LaAlO_3)_{0.3}(Sr_2TaAlO_6)_{0.7}$ (LSAT) (0 0 1) substrates. We estimate $\delta$ to be close to 0.5 based on the lattice constant and X-ray absorption spectroscopy (XAS) data (discussed later). During growth, the substrate heater was maintained at 600°C while the main chamber pressure was held at ~$2.5\times10^{-6}$ Torr after introducing $O_2$. Reflection high-energy electron diffraction (RHEED) was used to monitor the deposition *in situ*. The cation fluxes were evaporated from heated Sr and Mn metal sources and were measured by a quartz crystal monitor to determine the shuttering times. The metal cations were co-deposited for ~30 s per unit cell followed by a 10 s anneal pause. The atomic composition was calibrated by Rutherford backscattering spectroscopy (RBS) and X-ray photoelectron spectroscopy (XPS) depth profile. The total thickness of the as-grown SMO films is approximately 80 unit cells (~ 30 nm).

Before fluorination, each $10\times10$ mm$^2$ as-grown SMO film was cut with a sectioning saw into 9 equal square pieces of ~$3.3\times3.3$ mm$^2$. Each piece was then fluorinated with different fluoropolymers so that when comparing among the fluorinated films, the growth difference of the as-grown films is minimized. The fluorination method used was a vapor transport process, based on that reported by Katayama and coauthors [19]. In this approach, both the SMO film and fluoropolymer were placed in an alumina boat, separated by a piece of aluminum foil to prevent their physical contact. The boat was covered with aluminum foil to encourage the released F to react with the SMO film; then the boat was placed inside a quartz tube in a tube furnace. A schematic of the growth apparatus is shown in the Supplemental Material (Figure S1) [23]. The fluoropolymer was located upstream of the SMO film in flowing Ar gas with a 0.25 L/min flow rate. The mass of the polymer source used in each reaction was 0.5 g. The chemical formulas and decomposition temperatures of these polymers are listed in Table 1. One series of fluorination experiments was performed at 225°C for 30 min, another was at 235°C for 30 min. These reaction temperatures are lower than the polymer decomposition temperatures, although it should be noted that decomposition begins at temperatures lower than those listed in Table 1. For example, although the decomposition temperature of PTFE is 460°C [24], slight decomposition was observed as low as to 230°C [25]; this reduced initiation temperature for decomposition is observed in PVF and PVDF [24]. We choose fluorination temperatures of 225°C and 235°C to preserve the crystallinity of the fluorinated oxyfluoride films and to prevent formation of secondary phases [19].

Table 1. Chemical formulas and decomposition temperatures of PE, PTFE, PVF and PVDF [23,26-28].

| Polymer | PE | PTFE | PVF | PVDF |
|---|---|---|---|---|
| **Formula** | $(C_2H_4)_n$ | $(C_2F_4)_n$ | $(C_2H_3F)_n$ | $(C_2H_2F_2)_n$ |
| **Decomposition Temp. (°C)** | 370 | 460 | 240 | 350 |

X-ray diffraction (XRD) and X-ray reflectivity (XRR) were measured with a Rigaku SmartLab diffractometer and simulated and fitted, respectively, with GenX [29]. XRD 2θ-ω scans were performed around the 0 0 2 Bragg peak. X-ray reflectivity is a low angle scattering technique used to probe thin films, in which oscillations in the intensity arise from scattering contrast between the film and substrate. By fitting the reflectivity, quantities such as thickness, surface roughness, and density uniformity throughout the film can be quantified [30]. XPS was used to track the atomic composition depth profile through the entire film with the $Ar^+$ ion gun sputtering at 500 eV. The XPS depth profile and photoelectron spectra of each element was analyzed with CasaXPS [31]. To confirm the presence of F in the SMOF films and investigate the nominal oxidization state of Mn before and after fluorination and PE anneal, XAS was performed at the Advanced Light Source, Lawrence Berkeley National Laboratory, in electron yield (EY) and luminescence yield (LY) modes to probe the Mn *L*-edge and F *K*-edge at 300 K.

Density functional theory (DFT) calculations were performed within the spin-polarized generalized gradient approximation (GGA) with the plus Hubbard *U* correction [32] using the Perdew-Burke-Ernzerof (PBE) functional [33] as implemented in the Vienna Ab-initio Simulation Package (VASP) [34,35]. Projector-augmented wave (PAW) potentials [36] were used to describe the electron core-valence interactions with the following configurations: Sr ($4s^24p^65s^2$), Mn ($3p^64s^23d^6$), O($2s^22p^4$), and F($2s^22p^5$). The kinetic energy cutoff of the plane-wave basis set was 550 eV and a 4×4×6 and 8×8×6 Monkhorst-Pack grid [37] was used for the *k*-space sampling of the $SrMnO_{2.5}$ supercell and bulk $SrMnO_3$ perovskite structure, respectively. Brillouin zone integrations were performed with the tetrahedron method. The cell volume and atomic positions were relaxed until the forces on each atom were less than 5 meV/Å$^{-1}$. A Hubbard *U* = 3 eV was applied to the Mn 3*d* orbitals. Spin order was set to G-type antiferromagnetic (AFM) order for $SrMnO_3$, and E-type spin order for oxygen deficient systems ($SrMnO_{2.5}$) [38]. Formation energies were calculated for structures with one F atom to identify the preferred substitution/insertion site as follows:

$$E_f(Sr_{16}Mn_{16}O_{39}F) = E(Sr_{16}Mn_{16}O_{39}F) - 16 \times E(SrO) - 7 \times E(Mn_2O_3) - 2 \times E(MnO) - \frac{1}{2}E(F_2)$$

$$E_f(Sr_{16}Mn_{16}O_{40}F) = E(Sr_{16}Mn_{16}O_{40}F) - 16 \times E(SrO) - 8 \times E(Mn_2O_3) - 0 \times E(MnO) - \frac{1}{2}E(F_2).$$

## III. RESULTS AND DISCUSSION

Figure 1(a) presents the XRD data of as-grown SMO, PE annealed and PTFE, PVF, PVDF fluorinated (235°C for 30 min) SMOF films and Figure 1(b) shows the XRR data of these films. The as-grown sample was not exposed to a post-growth anneal. The obtained $c$-axis parameters from Figure 1(a) are marked as green circles in Figure 1(c) and the values are listed in Table 2. An additional set of $c$-axis parameters, measured from different films, of PE and fluoropolymers annealed films at 235°C for 30 min are plotted in yellow squares to illustrate the reproducibility of the results. For comparison, the $c$-axis parameters of those annealed at 225°C for 30 min films are plotted in blue circles. The average $c$-axis parameter of the as-grown films is 3.790 Å and is indicated with a red square where the error bar is the standard deviation among these three as-grown SMO films. The crystallinity of the epitaxial film can be described by the coherence length ($\xi$) derived from the full width at half-maximum (FWHM) ($\Delta q$) of the Bragg peak using $\xi = 2\pi/\Delta q$ [39]. Figure 1(d) shows the coherence length, normalized by that of the as-grown film, of the samples.

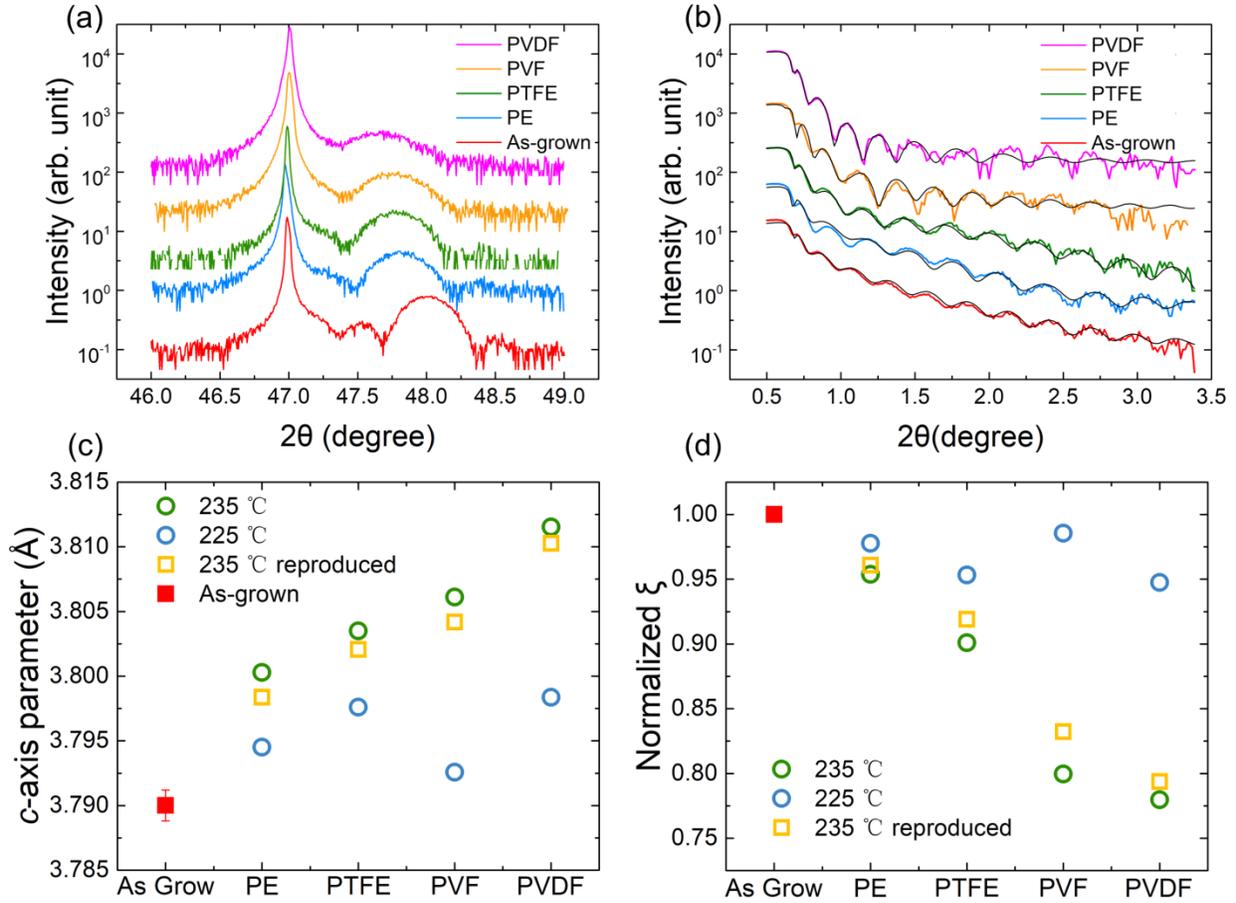

FIG. 1. (a) XRD data and (b) XRR data of as-grown, PE annealed and PTFE, PVF, PVDF fluorinated films at 235°C for 30 min. (c) The *c*-axis parameters of as-grown, PE annealed and fluorinated films at 225°C and 235°C for 30 min. (d) Normalized coherence length of the corresponding films in (c). Data represented by the green and blue circles are pieces from the same growth.

As shown in Figure 1(c), the *c*-axis parameter expands after reacting with both PE and fluoropolymers (PTFE, PVF, PVDF), but the lattice expansion is more pronounced after fluorination. For the 235°C reacted films, the lattice expansion is accompanied by a degradation of crystallinity, which is also reflected in the XRR data in Figure 1(b). The XRR data are plotted with the fitted curve shown in black. To accurately capture the XRR data, the PVF and PVDF fluorinated films had to be simulated using a model in which the film is divided into three layers and the PTFE fluorinated film is divided into two layers. The need to split the films into multiple layers suggests non-uniformity through the depth of the film. For the as-grown and PE annealed films, a single-layer model provides an optimal fit. Additionally, an increase in surface roughness ($R$) is observed in the PVDF ($R$ = 37 Å) and PVF ($R$ = 30 Å) samples, compared to the as-grown, PE- and PTFE-treated films ($R$ = 6-9 Å). The parameters of the XRR fitted data are listed in Table S1 of the supplemental material. The results are attributed to an inhomogeneous depth distribution of F in PVF and PVDF fluorinated films, a hypothesis supported by XPS.

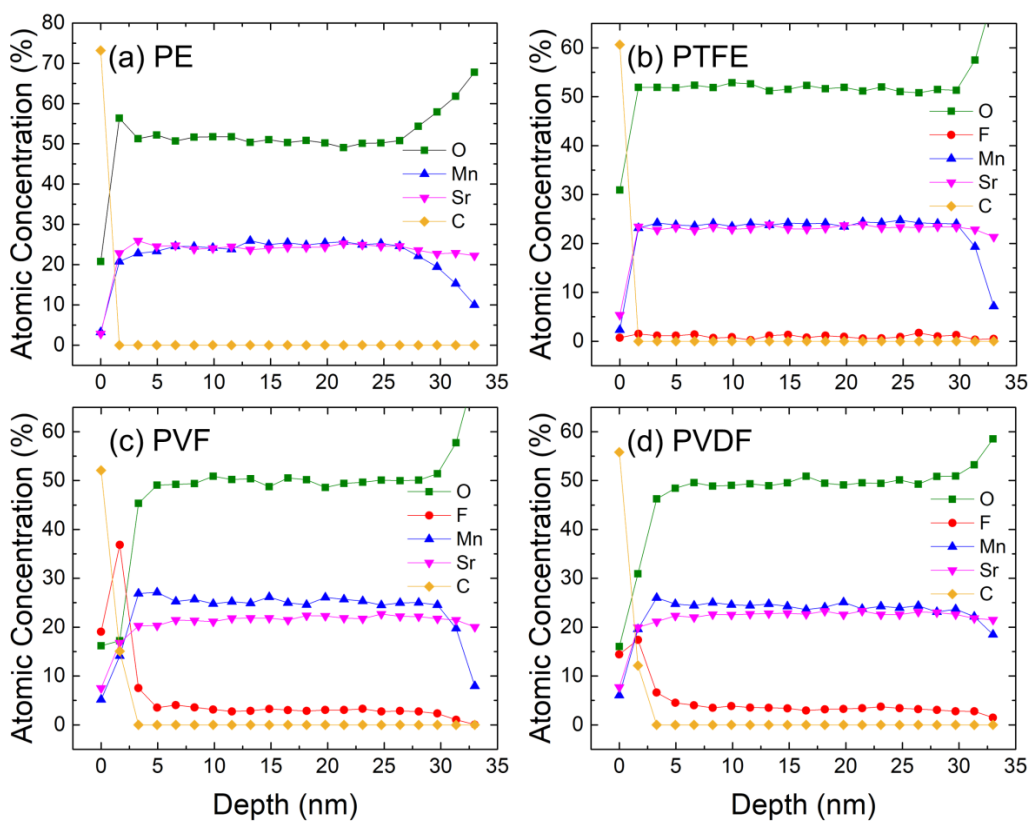

FIG. 2. XPS depth profiles show the atomic concentration at different sputtering depths of (a) PE annealed film, and (b) PTFE, (c) PVF, (d) PVDF fluorinated SMOF films at 235°C for 30 min. These data were measured from the samples denoted by yellow squares in Figure 1(c) and (d).

To probe the depth profile of the F concentration, XPS measurements were performed. Photoemission spectra were collected after 20 cycles of Ar$^+$ ion gun sputtering. After each cycle, the Sr 3$d$, Mn 2$p_{3/2}$, O 1$s$, and F 1$s$ photoelectron spectra were obtained. The data shown at "0-minute" sputtering time indicates the measurement from the surface prior to any sputtering; the depth profile reaches the LSAT substrate after 20 cycles. In converting from sputtering time to depth, we assume that the sputtering rate is uniform throughout the film and that each sputtering cycle removes 1.6 nm of material. Figure 2(a) shows the depth profile of the PE annealed film. There is a high C residue on the surface which is 73 % of the total atomic concentration compared with the 33 % of surface C on the as-grown SMO film, as shown in supplemental material Figs. S2 and S3. The increased amount of surface C residue also shows up in the fluorinated SMOF films but to a lesser degree than that of the PE annealed film. The XPS depth profiles of three SMOF films fluorinated at 235°C with PTFE, PVF, and PVDF are shown in Figure 2(b), (c), and (d), respectively. The PTFE fluorinated SMOF film has the least amount of F incorporated. PVF and PVDF produce higher F contents in the film; however, they both result in a high near-surface concentration of F, which gradually decreases over the first few sputtering cycles. The presence of a non-uniform F concentration near the surface is consistent with the structural models obtained from the XRR data (Table S1). In contrast, PTFE results in a more uniform distribution of F through the film. Even after exposure to longer reaction times with PTFE to increase the F content in the film, the near-surface region of the PTFE-reacted SMOF film has significantly less F accumulation compared to the PVDF and PVF-reacted films, as shown in Fig. S4. The average F concentration (γ) is obtained by averaging over cycles 5 to 15 and normalizing by the average of Sr and Mn composition. The precursor fluoropolymer, $c$-axis parameter, and γ value with standard deviation of each film are listed in Table 2.

Table 2. F contents and *c*-axis parameters of as-grown, PE annealed, and films fluorinated at 235°C for 30 minutes. The results were obtained from the films marked as green circles and yellow squares in Figure 1(c).

| **Film** | **As-grown** | **Annealed** | **Fluorinated $SrMnO_{3-\delta}F_\gamma$** | | |
|---|---|---|---|---|---|
| **Polymer** | None | PE | PTFE | PVF | PVDF |
| ***c*-axis (Å)** | 3.790±0.02 | 3.799±0.01 | 3.803±0.01 | 3.805±0.01 | 3.811±0.01 |
| **F content ($\gamma$)** | 0 | 0 | 0.04±0.015 | 0.15±0.013 | 0.13±0.015 |

As expected, the PVF and PVDF fluorinated SMOF films, which show larger lattice expansions, have higher F content than the PTFE fluorinated film. This fluorination-induced lattice expansion has been reported in bulk $SrMnO_{2.5-x}F_{0.5+x}$ and $SrFeO_{3-\delta}F_\gamma$ oxyfluorides [9,13,19]. In the oxyfluoride SMOF films, there are both Mn-F and Mn-O bonds, and the valence state of Mn cation could be either reduced or oxidized after fluorination depending on whether the F substitutes for O or is inserted into the anion vacancy sites.

DFT calculations were performed to evaluate the formation energy and structural changes upon fluoride insertion into $SrMnO_{3-\delta}$. In the calculations, the $SrMnO_{2.5}$ structure was assumed to exhibit the oxygen vacancy ordered pattern found in bulk $Sr_2Mn_2O_5$ [38], using a 2×1×2 supercell of the primitive $SrMnO_{2.5}$ unit cell to accommodate and approximate the experimental fluoride concentration. Three $Sr_{16}Mn_{16}O_{39}F$ (equivalent to $SrMnO_{2.4375}F_{0.0625}$) structures were generated with F substituting on different oxygen sites (O1, O2 and O3), and $Sr_{16}Mn_{16}O_{40}F$ ($SrMnO_{2.5}F_{0.0625}$) was constructed with F occupying a vacancy site, as shown in Figure 3(a). Models consisting of two fluorine atoms were generated based on oxygen substitution and at different distances; one with F located on the same polyhedral unit and another with F located further away in the supercell.

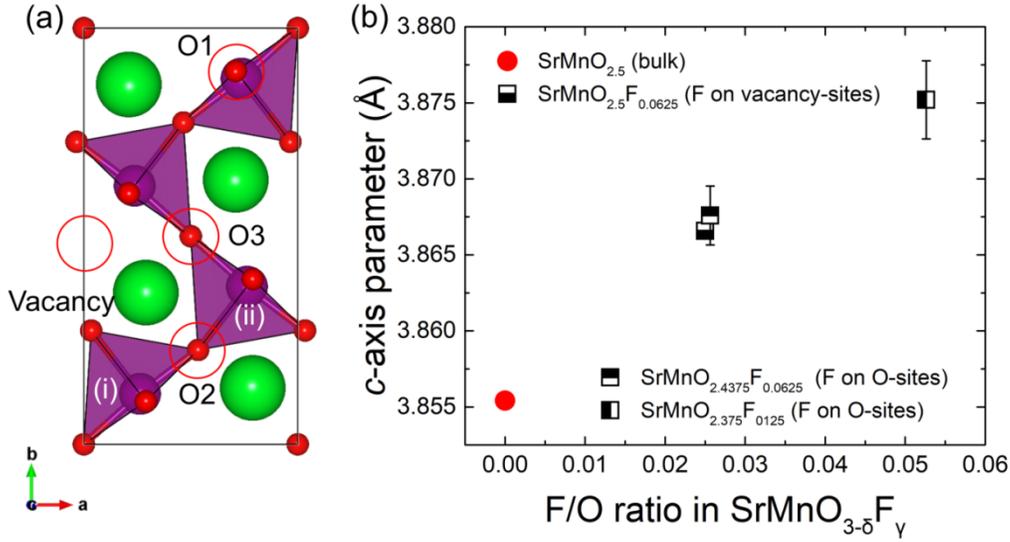

FIG. 3. (a) SrMnO$_{2.5}$ structure and different O and vacancy sites. The purple polyhedral units are the MnO$_5$ square pyramids. The black box denotes the Sr$_2$Mn$_2$O$_5$ unit cell. The red circles highlight different anion sites within the structure. The circle on the left edge of the unit cell indicates a vacancy site, while the three circles within the unit cell indicate the O1, O2, and O3 sites. (b) DFT calculated $c$-axis parameter as a function of F/O ratio in SrMnO$_{3-\delta}$F$_\gamma$.

Table 3. The DFT calculated average Mn-F bond length in Å. For the SrMnO$_{2.5}$ case, the provided value corresponds to the Mn-O bond length. For comparison, the average bond length within the MnO$_6$ octahedra in perovskite SrMnO$_3$ is 1.923 Å.

| System | SrMnO$_{2.5}$ | F on O1-site | F on O2-site | F on O3-site | F on vacancy-site |
|---|---|---|---|---|---|
| Ave. Bond Length | 1.971 | 2.006 | 2.063/1.974 | 2.014 | 1.966 |

From the DFT calculations, the $c$-axis parameter generally increases with fluorination, as shown in Figure 3(b). We find an almost linear trend of the $c$-axis parameter with fluorine incorporation into the oxide. When F occupies the vacant anion site, the cell volume increases owing to the change in occupied space. Three different O sites are possible for exchange as shown in Figure 3(a) and the corresponding Mn-F bond length for each substitution is listed in Table 3. Fluorine substitution on the O1 and O3 sites corresponds to replacing the equatorial positions for both neighboring Mn atoms, whereas the O2-site substitution produces both an apical and equatorial Mn-F bond, yielding two different bond lengths with an average value larger than F substitution on the equatorial sites. When F occupies the vacancy site, the average bond length in the $MnO_5F_1$ octahedra decreases despite the oxyfluoride octahedra having longer equatorial Mn-O bonds than $SrMnO_{2.5}$ and longer average bond lengths than that of octahedral units in perovskite $SrMnO_3$. Thus, the Mn-F bonds in all situations are longer than the average Mn-O bond length in $SrMnO_{2.5}$ or $SrMnO_3$ without fluorination, which indicates that the formation of Mn-F bonds in the produced oxyfluoride SMOF films contributes to the lattice expansion.

When $F^-$ substitutes on an $O^{2-}$ site, reduction of the coordinating transition metal is required. Chemically, the different anion charges reduce $Mn^{3+}$ to $Mn^{2+}$, which yields a higher ionic radius for the lower oxidation state metal. This chemical expansivity effect also explains the comparably smaller $c$-axis parameter found in bulk $SrMnO_3$ that consists of Mn in a 4+ oxidation state, i.e., exhibiting an ionic radius that is smaller than both $Mn^{3+}$ and $Mn^{2+}$. The reduction of Mn with fluorine substitution is further supported by our DFT analysis of the local Mn magnetic moments. Although the DFT-calculated magnetic moment of Mn in bulk $SrMnO_{2.5}$ is 3.7 $\mu_B$, the fluorinated systems exhibit calculated moments greater than 4 $\mu_B$ on the Mn cations coordinated by F. Thus, the reduction of Mn by fluorine increases the $c$-axis parameter of the fluorinated materials.

Interestingly, the Mn reduction is found to be fluorine site dependent. For O1- and O3-site substitutions, the magnetic moment of two Mn atoms connected by the bridging fluoride anion exhibit local moments of 4.03 $\mu_B$ and 4.12 $\mu_B$ respectively. For O2-site substitution, on the other hand, the coordinating Mn cations become distinct with 4.39 $\mu_B$ and 3.76 $\mu_B$ for the (i) and (ii) square pyramids shown in Figure 3(a). Noting that the calculated magnetic moment of bulk $SrMnO_{2.5}$ is 3.7 $\mu_B$, it seems that the excess electron provided by fluorination is strongly tied to the square pyramidal Mn (i), while the other Mn atom on square pyramid (ii) remains largely unchanged. The change in the average Mn-O/F bond length ($l_{ave}$) of square pyramids is consistent with our interpretation of the magnetic moment [(Figure 3(b)]. While $l_{ave}$ of the square pyramids with O1- and O3-site substitution are similarly longer than that of bulk $SrMnO_{2.5}$, square pyramid (i) has the longest $l_{ave}$, and square pyramid (ii) has an $l_{ave}$ value similar to the bulk $SrMnO_{2.5}$ value.

Further evidence of Mn-F bonds comes from the Mn $2p_{3/2}$ photoelectron peak position obtained by XPS. Figure 4 shows the Mn $2p$ photoelectron peaks of as-grown, PE annealed and fluorinated films. The spectra are obtained by averaging the measured spectra from cycle 5 to 15. The Mn $2p_{3/2}$ peak shifts to higher binding energy after fluorination but exhibits minimal change after the PE anneal. Previous studies have reported the Mn $2p_{3/2}$ peak position at 641.2-641.7 eV in MnO and 641.4-641.8 eV in $Mn_2O_3$ [40,41]. The difference of the Mn $2p_{3/2}$ peak location between MnO and $Mn_2O_3$ is less than 0.2 eV. Therefore, the Mn $2p_{3/2}$ peak location does not provide insight into the Mn valence change after the PE anneal. As for manganese fluorides, the binding energy of the Mn $2p_{3/2}$ peak is situated at 642.8 eV for both $MnF_2$ and $MnF_3$ [41]. Our observation that the fluorinated films exhibit Mn $2p_{3/2}$ peak shifts to higher binding energy, is thus consistent with the formation of Mn-F bonds. Combining the DFT and XPS results, we conclude that the longer bond length of Mn-F compared to Mn-O justifies the fluorination-induced lattice expansion.

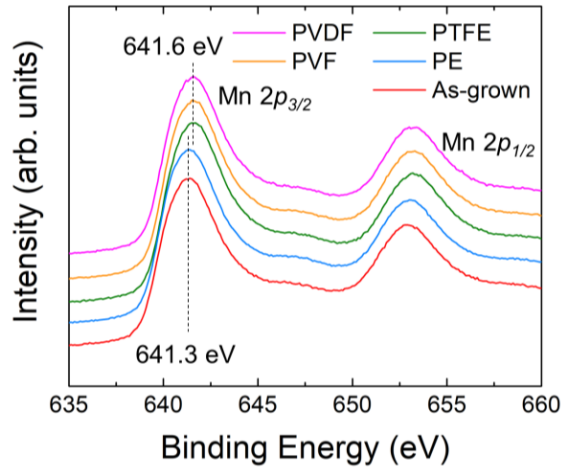

FIG. 4. The Mn 2*p* photoelectron spectra of different films from XPS measurements.

We note that the PE annealed film also shows a lattice expansion even though there is no F involved. Considering the ionic size of $Mn^{2+}$ is larger than $Mn^{3+}$, any reduction of the as-grown film will also result in lattice expansion [13,19]. Thus, we believe that the C and H species released from the decomposing polymers contribute to oxygen removal from the film.

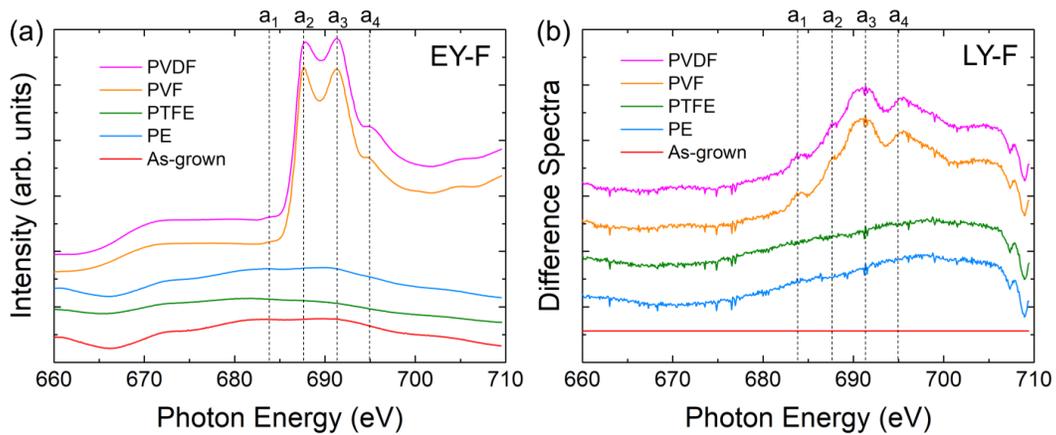

FIG. 5. F *K*-edge XAS spectra of as-grown SMO, PE annealed and fluorinated SMOF films. (a) The EY signal of the surface F. (b) The difference spectra of LY signal obtained by subtracting the spectrum of the as-grown SMO film.

XAS was measured near the Mn *L*-edge to determine the nominal valence states of Mn and also near the F *K*-edge. Figures 5 and 6 shows the F *K*-edge and Mn *L*-edge XAS spectra, respectively, for all five films for which the structural data is plotted in Figure 1(a,b). The F *K*-edge EY and LY signal were obtained over the photon energy range of 660 to 710 eV. The EY signal is acquired while monitoring the sample drain current, which is due to the emission of photoelectrons created by the absorbed X-rays. There is a sampling depth limitation of the EY signal, which makes it a surface sensitive mode [42]. In contrast, the LY detection mode is bulk sensitive, capturing information over the entire film thickness [43,44]. As shown in Figure 5, the F *K*-edge EY and LY spectra are observed only in the PVF and PVDF fluorinated SMOF films, consistent with the higher F concentration in PVF and PVDF fluorinated films probed by XPS in Figure 2. Figure 5(b) displays the F *K*-edge LY difference spectra, obtained by subtracting the spectrum of the as-grown film from the spectra obtained from films after reaction/annealing. The difference spectra exhibit four peaks near the F *K*-edge, labeled as $a_1$ to $a_4$ in Figure 5(a,b). The F spectra obtained by EY of PVF and PVDF fluorinated SMOF films [Fig. 5(a)] resemble those previously reported for $MnF_2$ [45,46], which shows a similar double-peak $a_2$ and $a_3$ feature. Meanwhile, the low-intensity $a_1$ peak located at lower binding energy than the $a_2$ peak is consistent with the lower energy position of the F *K*-edge in $MnF_3$ than that in $MnF_2$ [45], which indicates some amount of F bonded with $Mn^{3+}$. In the LY data, the $a_2$ and $a_3$ peaks become less intense while the $a_1$ peak intensity increases compared to the EY data. We conclude that on the surface, F ions are dominantly bonded with $Mn^{2+}$ but within the film there is a considerable amount of $Mn^{3+}$-F bonds.

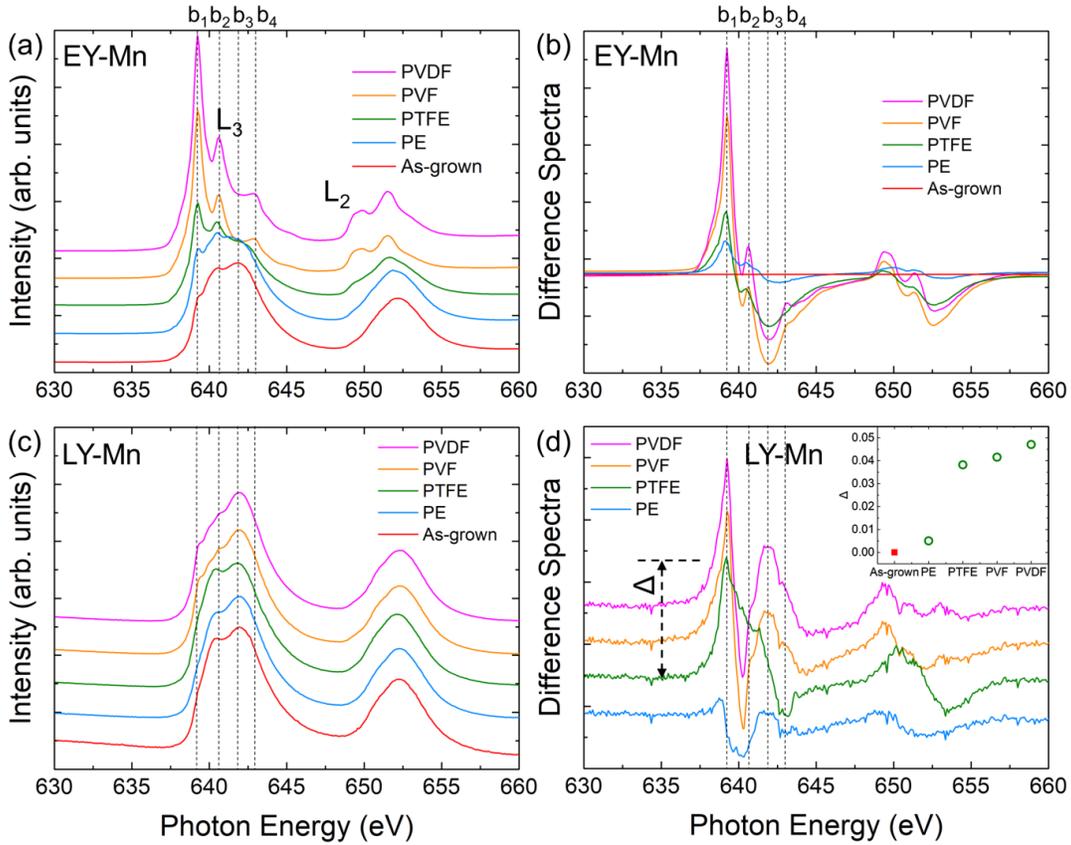

FIG. 6. Mn $L$-edge XAS spectra of as-grown SMO, PE annealed and fluorinated SMOF films. (a) The measured EY signal and (b) the difference spectra of the EY signal obtained through subtraction of the normalized spectrum of as-grown SMO film. (c) The measured LY signal and (d) the difference LY spectra. The inset shows the height ($\Delta$) of the peak located at 639.2 eV, obtained from the LY data.

To confirm the oxidation state of the surface and bulk Mn, the Mn $L$-edge XAS spectra were measured in both EY and LY mode, as shown in Figure 6. Previously, Mn $L$-edge XAS spectra of different Mn valence states have been measured from manganese oxide $MnO_x$ [47], manganese fluoride $MnF_x$ [45], and manganite perovskite $La_xMnO_{3-\delta}$ [48]. We use the Mn $L$-edge XAS spectra of these materials to analyze the spectra from the SMOF films. Additionally, the Mn $L$-edge XAS spectra for both $Mn^{2+}$ and $Mn^{3+}$ were simulated in CTM4XAS [49] and plotted in supplemental material (Figure S5) with the difference spectra obtained by subtracting the $Mn^{3+}$ from the $Mn^{2+}$ spectrum. The relevant simulation parameters are listed in supplemental material (Table S2) as well. The simulated spectra are used as a reference for our measured XAS data.

For the XAS data of the as-grown SMO film, EY and LY signals are shown in Figure 6(a) and (c). Both present the $Mn^{3+}$ feature that confirms the stoichiometry of our as-grown SMO film is very close to $SrMn^{3+}O_{2.5}$. The Mn *L*-edge EY signal of PVF and PVDF fluorinated films presented in Figure 6(a) is similar to the $Mn^{2+}$ spectra from previous experimental reports [45,47,50] and the simulation in Figure S5. Therefore, the surface Mn of PVF and PVDF fluorinated films are dominated by $Mn^{2+}$, which is consistent with the intense $a_2$ and $a_3$ peaks in Figure 5(a) of these SMOF films originating from $Mn^{2+}$-F bonds.

There are four peaks marked in the Mn $L_3$-edge spectra in each panel of Figure 6. These spectral peaks $b_1$ to $b_4$ are located at 639.2 eV, 640.5 eV, 641.8 eV and 642.9 eV, which are comparable with the four peaks at 640.4 eV, 642.0 eV, 642.8 eV and 644.0 eV associated with $Mn^{2+}$, $Mn^{3+}$ and $Mn^{4+}$ oxidation states in $MnO_x$ [46,47,48,50]. Among them, feature $b_1$ (located at 639.2 eV in our SMOF films, 640.4 eV in MnO [50], and 640 eV in $MnF_2$ [45]) is assigned to $Mn^{2+}$. These four measured peaks $b_1$ to $b_4$ match up well with the simulated spectra in Figure S5.

The difference spectra of the Mn *L*-edge in Figure 6(b,d) were obtained by subtracting the spectrum of the as-grown film from those of the reacted films. The intensity of peak $b_1$ in the difference spectra, originating from $Mn^{2+}$ and denoted by Δ, becomes larger in both the EY and LY signals after the PE anneal and fluorination, but the effect is especially pronounced after fluorination. As shown in the inset of Figure 6(d), there is an abrupt increase of Δ between the fluorinated films and PE annealed film, indicating that the $Mn^{2+}$ component increased significantly after fluorination. In other words, fluorination reduces the Mn from the as-grown SMO film to a greater extent than the PE anneal.

The Mn oxidization states before and after fluorination provide insight into the F incorporation mechanism of the vapor transport fluorination process. There are three ways for F to incorporate into oxygen deficient perovskite compounds: F occupation at an oxygen vacancy, F substitution of an in-site oxygen, and F insertion into an interstitial site [52,53]. The F interstitial insertion usually happens in layered perovskite compounds accompanied by a large expansion of 1 ~ 3 Å along the *c*-axis [54,55], which is ten times more than what we observe in our SMOF films. Thus, the possible ways for F incorporation in our films are F/O-vacancy occupation or F/O substitution. For $\gamma \neq 0$ in $SrMnO_{3-\delta}F_{\gamma}$, F/O-vacancy occupation will oxidize the $Mn^{3+}O_5$ pyramids in as-grown film into $Mn^{3.5+}O_5F$ octahedra. In contrast, the F/O substitution reaction will reduce the $Mn^{3+}O_5$ pyramids into $Mn^{2.5+}O_4F$. According to the XAS analysis, the films were reduced after fluorination, which supports the F/O substitution mechanism and is consistent with the previously discussed DFT results in which F substitution reduces the coordinated Mn cation. Consequently, the more F is incorporated, the more O is being substituted which will result in further reduction of Mn. Comparing the F content probed by XPS in Figure 2 with the difference spectra in Figure 6, the PVF and PVDF fluorinated films have higher $\gamma$ and a larger $Mn^{2+}$ feature compared with the PTFE fluorinated film.

This experimental evidence for fluorine substitution reactions, instead of fluorine insertion, is consistent with formation energies obtained from DFT, listed in Table 4. Among the F/O substitutions, the F substitution on the O2-site has the lowest formation energy of -13.26 eV. Meanwhile, the F/O-vacancy occupation has a formation energy of -13.22 eV, which is slightly higher than the F/O2 substitution. These formation energy trends are maintained with various density functionals with our PBE+U=3eV results shown here providing close to a lower limit on the energetic site-occupancy stability. Therefore, F substitution on the O2-site is the most energetically favorable fluorination pathway explored, which supports the F/O substitution mechanism. Considering our computed formation energies in Table 4, it is noteworthy that F substitution on the O2-site is distinctly more stable than other substitution sites. The O2-site corresponds to the equatorial anion for one pyramidal unit and the apical anion for another square pyramidal unit, whereas the O1- and O3-sites are always equatorial anions. The square pyramid having F on an equatorial anion site more strongly affects the local $d$-orbital configuration for that ion than when the F is on apical site. The resulting asymmetric reduction of Mn atoms ($Mn^{2+}$ and $Mn^{3+}$) is energetically more stable than at substitution sites which produce pairs of $Mn^{3.5+}$.

Table 4. The formation energies calculated from DFT of different fluorine occupations.

| Site occupation | Composition | Formation Energy (eV/supercell) |
|---|---|---|
| F on O1-site | $Sr_{16}Mn_{16}O_{39}F$ | -12.93 |
| F on O2-site | $Sr_{16}Mn_{16}O_{39}F$ | -13.26 |
| F on O3-site | $Sr_{16}Mn_{16}O_{39}F$ | -12.92 |
| F on vacancy site | $Sr_{16}Mn_{16}O_{40}F$ | -13.22 |

For the PE annealed film, there is no F vapor, but the manganite is still being reduced. In view of this, besides the F/O substitution induced film reduction, C and/or H in the polymers also acts as a reducing agent, a result previously reported in bulk metal oxides [22]. The deconvoluted surface C 1$s$ XPS spectra are shown in Figure S2 to study the C components. The PE annealed film has the highest concentration of surface C residue in the form of C-C/C-H bonds, which indicates the surface residues are mainly hydrocarbon materials. In contrast, there are more C to O bonds on the surface of fluorinated SMOF films as shown in S2(b)-(d), and the percentage of C-O/O-C=O bonds increases in the order of PTFE, PVDF, and PVF fluorinated films as shown in Figure S3. In addition, there is no C-F peak shown in the C 1$s$ spectra [56-58], which suggest the surface C residue are from C-O/O-C=O or C-C/C-H instead of C-F. The reduction level of $Mn^{3+}$ after fluorination is correlated with the percentage of surface C-O/O-C=O bonds on SMOF films, especially the PVF and PVDF fluorinated films. This further suggests that the C in polymers acts as a reducing agent by bonding with O in the SMO film, while F reduces the as-grown film by substituting for the in-site O as previously discussed.

## IV. CONCLUSION

We synthesized epitaxial $SrMnO_{3-\delta}F_\gamma$ ($\gamma < 0.15$) oxyfluoride films via a topochemical vapor transport method using PTFE, PVF, and PVDF as fluorinating agents. PVF and PVDF result in a higher amount of incorporated F and more crystalline degradation. By comparison, there is only a small amount of F incorporated after fluorination with PTFE for 30 minutes but it better preserves crystallinity and has a more uniform F distribution without the high surface F residue. The Mn 2$p_{3/2}$ photoelectron peak shifts to higher binding energy after fluorination indicative of the formation of Mn-F bonds, while DFT calculations point to the formation of Mn-F bonds as the origin of the observed fluorination-induced lattice expansion. The Mn XAS spectra reveal that the as-grown films are reduced after fluorination, which supports the F/O substitution mechanism as opposed to simple F insertion into anion vacancies. The DFT calculations confirm that the formation energy of F substituting on the O2-site is thermodynamically preferred over that of F insertion into an O-vacancy site. Even though there is no F in PE, annealing with PE also slightly reduces the films suggesting that the C vapor acts to remove oxygen from the films. Therefore, we propose that the fluorination of $SrMnO_{2.5}$ films via vapor transport from fluoropolymers involves F/O substitution and C/O bonding, both of which reduce the as-grown manganites leading to mixed $Mn^{3+}/Mn^{2+}$ in the fluorinated films.

**ACKNOWLEDGMENTS**

Work at Drexel was supported by the National Science Foundation (NSF, grant number CMMI-1562223). Film synthesis utilized deposition instrumentation acquired through an Army Research Office DURIP grant (W911NF-14-1-0493). Y.S and J.M.R. were supported by NSF (DMR-1454688). This research used resources of the Advanced Light Source, which is a DOE Office of Science User Facility under contract no. DE-AC02-05CH11231.

Supplemental Material for:

# Effect of Fluoropolymer Composition on Topochemical Synthesis of $SrMnO_{3-\delta}F_{\gamma}$ Oxyfluoride Films

Jiayi Wang,[1] Yongjin Shin,[2] Elke Arenholz,[3] Benjamin M. Lefler,[1] James M. Rondinelli,[2] Steven J. May[1]

[1]*Department of Materials Science and Engineering, Drexel University, Philadelphia, Pennsylvania 19104, USA*

[2]*Department of Materials Science and Engineering, Northwestern University, Evanston, Illinois 60208, USA*

[3]*Advanced Light Source, Lawrence Berkeley National Laboratory, One Cyclotron Road, Berkeley, California 94720, USA*

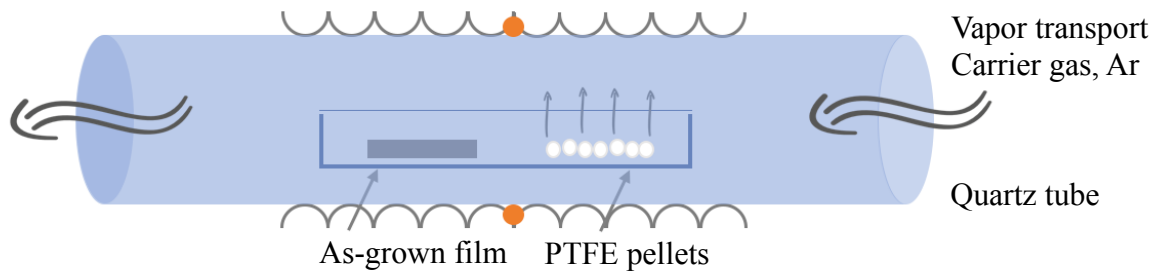

Figure S1. Schematic of the vapor transport fluorination process. The as-grown SMO film is placed downstream of fluoropolymer pellets in an alumina boat, separated by a barrier made of aluminum foil to prevent their physical contact. The whole alumina boat is wrapped in aluminum foil with two small holes punctured on both end to preserve the fluorine containing vapor products near the sample.

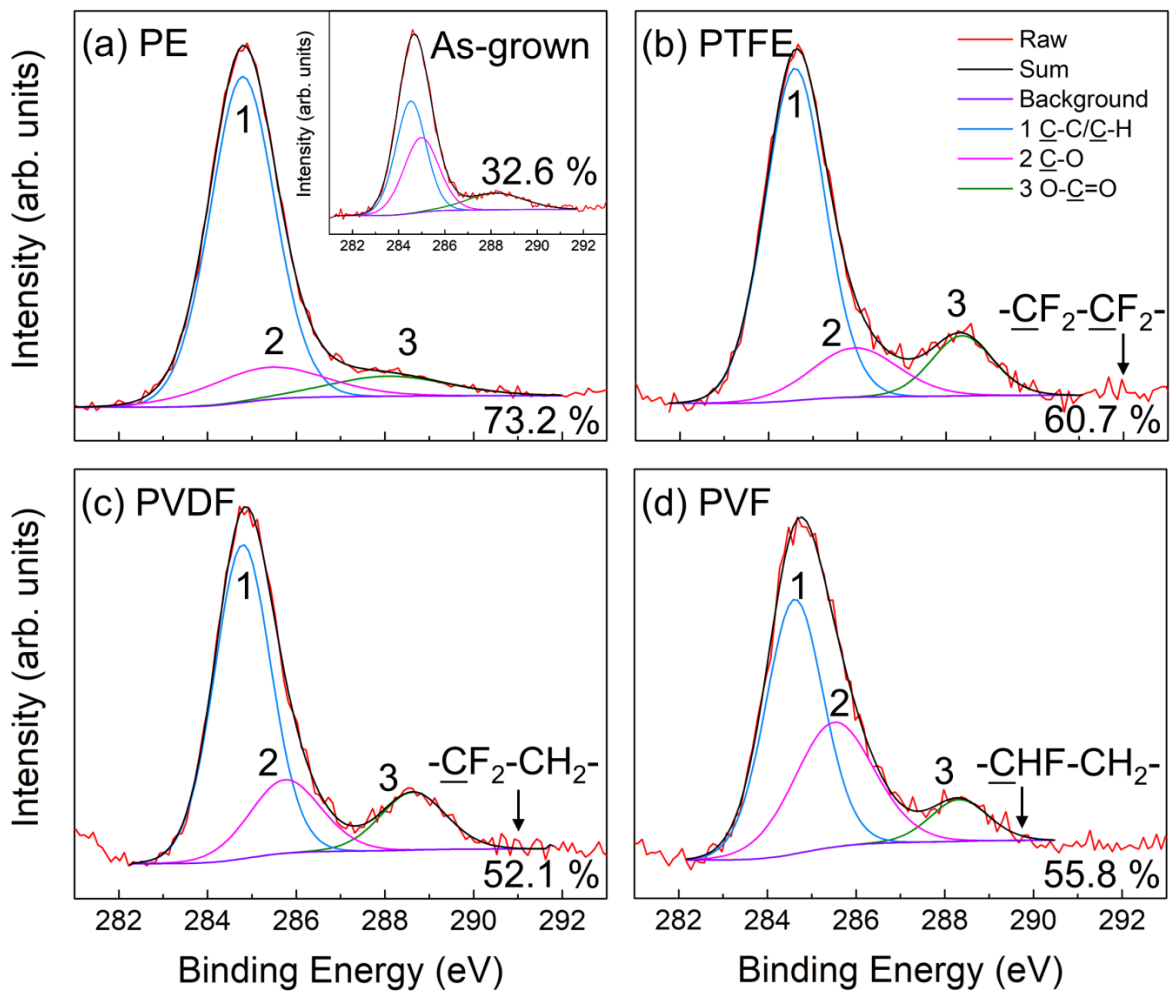

Figure S2. The surface C 1$s$ XPS spectra of (a) PE annealed film with inset of as-grown SMO film, (b) PTFE-, (c) PVF-, and (d) PVDF-fluorinated SMOF films. Each spectrum is deconvoluted into three components representing the C in different chemical environments. The percentage numbers at the bottom right corner of each figure is the surface atomic concentration of C. The typical C-F peak position of each fluoropolymer is marked but experimentally this peak is not observed.

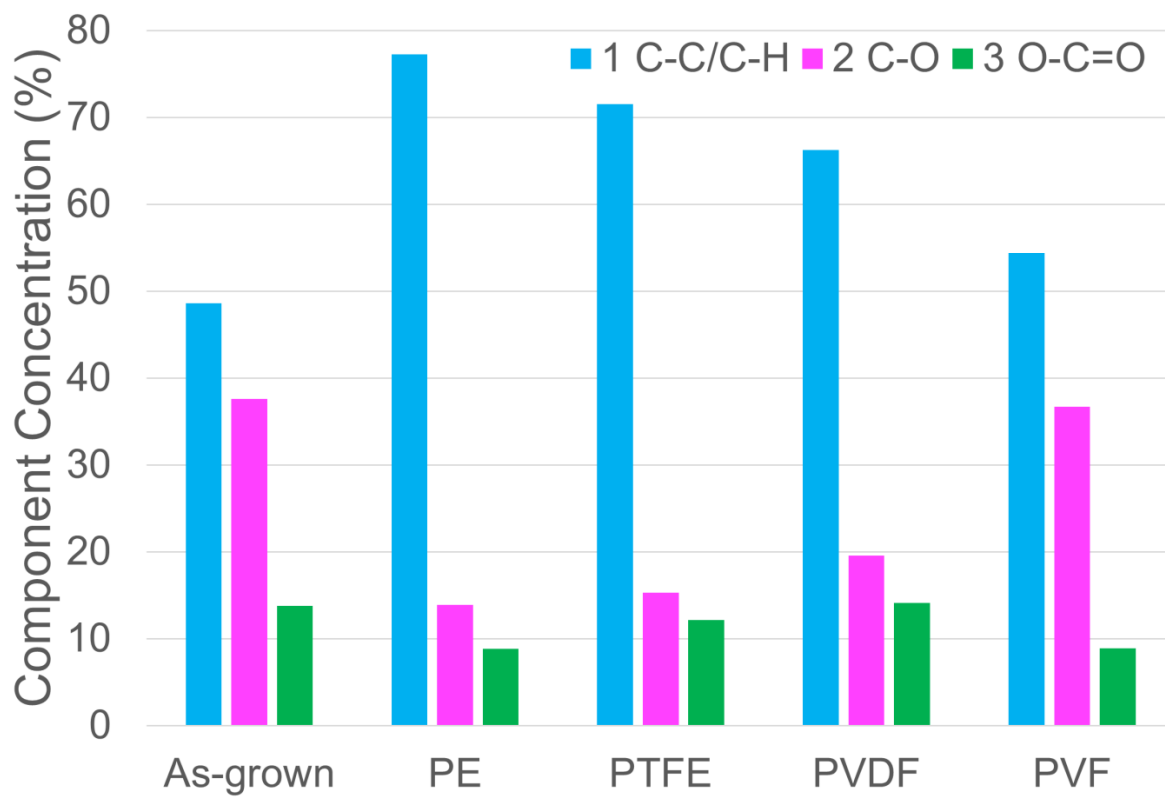

Figure S3. The percentage of C-C/C-H, C-O, and O-C=O components of the surface C residue on the surface of each film, which are obtained from the surface C 1*s* XPS spectra in Figure S2.

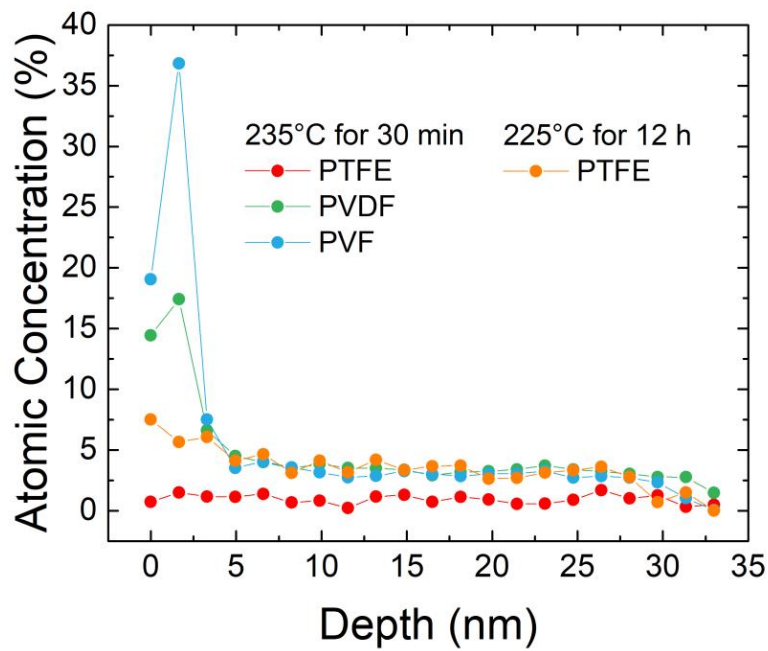

Figure S4. The atomic concentration of F in SMO films. After exposure for 12 hours, the PTFE-reacted film has a similar F concentration as the films reacted with PVF and PVDF for 30 minutes but with a significantly reduced surface F concentration.

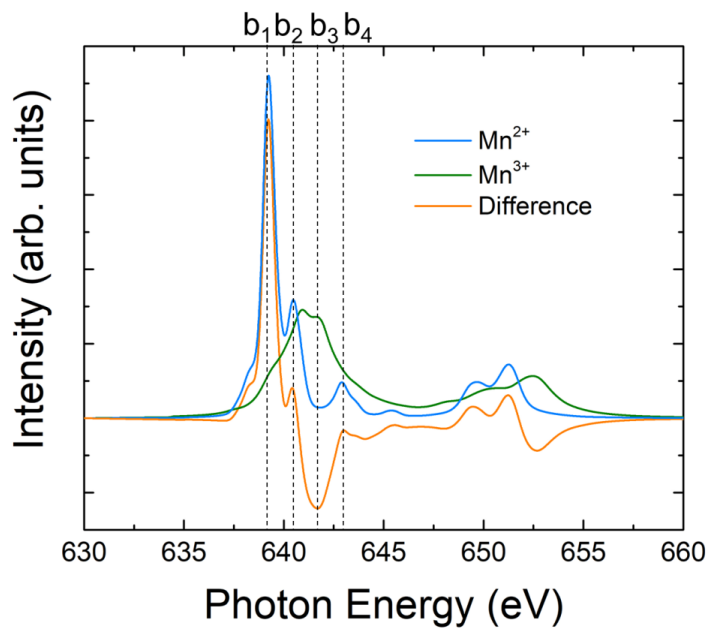

Figure S5. The Mn *L*-edge XAS spectra simulation of $Mn^{2+}$ and $Mn^{3+}$.

Table S1. The parameters obtained from the fits to XRR data of each film.

| Film | | SMO | | SMOF | | |
|---|---|---|---|---|---|---|
| Reacting polymer | | None | PE | PTFE | PVF | PVDF |
| Surface roughness (Å) | | 8 | 9 | 6 | 30 | 37 |
| Thickness (Å) | Layer 1 | 303 | 268 | 39 | 53 | 102 |
| | Layer 2 | | | 238 | 217 | 160 |
| | Layer 3 | | | | 20 | 61 |

Table S2. Simulation parameters of $Mn^{2+}$ and $Mn^{3+}$ $2p$ spectra in CTM4XAS.

| **Electronic configuration** | | **$Mn^{2+}$** | **$Mn^{3+}$** |
|---|---|---|---|
| **Slater integral reduction (%)** | $F_{dd}$ | 1 | 1 |
| | $F_{pd}$ | 1 | 1 |
| | $G_{pd}$ | 1 | 1 |
| **SO coupling reduction (%)** | Core | 1 | 1 |
| | Valence | 1 | 0 |
| **Crystal field parameters (eV)** | Symmetry | $O_h$ | $O_h$ |
| | 10 $D_q$ initial | 0.6 | 0.9 |
| | 10 $D_q$ final | 0.6 | 0.9 |
| **Charge transfer parameters (eV)** | Delta | 2 | 1 |
| | $U_{dd}$ | 1 | 1 |
| | $U_{pd}$ | 0 | 0 |
| **Plotting** | Lorentzian broadening | 0.1/0.4 | 0.5/0.7 |
| | Split | 647 | 648 |
| | Gaussian broadening | 0.2 | 0.1 |
| | Temperature (K) | 300 | 300 |

**Formation Energy via Density Functional Theory Calculation**

The formation energy of fluorinated $SrMnO_{2.5}$ was calculated with generalized gradient approximation (GGA) with the Hubbard *U* correction using the Perdew-Burke-Ernzerof (PBE) functional. To accommodate and approximate the experimental fluoride concentration, a 2×1×2 supercell of the primitive $SrMnO_{2.5}$ unit cell was used. There are multiple ways to define the formation energies using different compounds of the (Sr – Mn – O – F) system as reference. We choose to use those compounds with low formation energies, $Mn_2O_3$ and MnO, based on the Open Quantum Materials Database (OQMD) [1]. With these states, the formation energy calculations for the oxygen-site substitution and F-insertion into vacancy-site were obtained as follows:

$$E_f(Sr_{16}Mn_{16}O_{39}F) = E(Sr_{16}Mn_{16}O_{39}F) - 16 \times E(SrO) - 7 \times E(Mn_2O_3) - 2 \times E(MnO) - \frac{1}{2}E(F_2)$$

$$E_f(Sr_{16}Mn_{16}O_{40}F) = E(Sr_{16}Mn_{16}O_{40}F) - 16 \times E(SrO) - 8 \times E(Mn_2O_3) - 0 \times E(MnO) - \frac{1}{2}E(F_2)$$

We also note that the formation energies can change depending on the choice of functionals in density functional calculations, and the PBE functional is known to provide most often the smallest formation energies [2,3]. Therefore, we deduced that the formation energy difference among different site-occupancy configurations will be larger when different functionals are applied, further supporting the assignment of the F-on-O2-site configuration.

Reference
[1] J. E. Saal, S. Kirklin, M. Aykol, B. Meredig, C. Wolverton *JOM* **65**, 1501 (2013)
[2] L. Schimka, J. Harl, and G. Kresse, *Journal of Chemical Physics* **134**, 024116 (2011)
[3] V. L. Chevrier, S. P. Ong, R. Armiento, M. K. Y. Chan, and G. Ceder, *Physical Review B* **82**, 075122 (2010)